\documentclass[11pt]{article}
\usepackage{amsmath,amssymb,color,graphics,cite,epsfig,comment}
\usepackage{hyperref}
\usepackage [latin1]{inputenc}
\usepackage{amsfonts}
\newcommand{\hoch}[1]{$\, ^{#1}$}

\newcommand{\be}{\begin{equation}}
\newcommand{\ee}{\end{equation}}
\newcommand{\bea}{\setlength\arraycolsep{2pt} \begin{eqnarray}}
\newcommand{\eea}{\end{eqnarray}}
\newcommand{\nn}{\nonumber}

\def\ft#1#2{{\textstyle{\frac{\scriptstyle #1}{\scriptstyle #2} } }}

\def\0{{\sst{(0)}}}
\def\1{{\sst{(1)}}}
\def\2{{\sst{(2)}}}
\def\3{{\sst{(3)}}}
\def\4{{\sst{(4)}}}
\def\5{{\sst{(5)}}}
\def\6{{\sst{(6)}}}
\def\7{{\sst{(7)}}}
\def\8{{\sst{(8)}}}
\def\sst#1{{\scriptscriptstyle #1}}

\def\d{\delta}

\def\k{\kappa}
\def\l{\lambda}

\def\m{\mu}
\def\n{\nu}
\def\r{\rho}
\def\s{\sigma}

\def\o{\omega}
\def\O{\Omega}

\newcommand{\w}[1]{\\[0.#1cm]}
\begin{document}

\begin{center}
{\Large {\bf Leading Higher Derivative Corrections to Multipole Moments of Kerr-Newman Black Hole }}

\vspace{20pt}

Liang Ma\hoch{1}, Yi Pang\hoch{1,3} and  H. L\"{u}\hoch{1,2,3}

\vspace{10pt}

{\it \hoch{1}Center for Joint Quantum Studies and Department of Physics,\\
School of Science, Tianjin University, Tianjin 300350, China }

\bigskip

{\it \hoch{2}Joint School of National University of Singapore and Tianjin University,\\
International Campus of Tianjin University, Binhai New City, Fuzhou 350207, China}

\bigskip

{\it \hoch{3}Peng Huanwu Center for Fundamental Theory,\\
Hefei, Anhui 230026, China}

\vspace{40pt}

\underline{ABSTRACT}
\end{center}

We study the (leading) 4-derivative corrections, including both parity even and odd terms, to electrically-charged Kerr-Newman black holes. The linear perturbative equations are then solved order by order in terms of two dimensionless rotating and charge parameters. The solution allows us to extract the multipole moments of mass and current from the metric as well as the electric and magnetic multipole moments from the Maxwell field. We find that all the multipole moments are invariant under the field redefinition, indicating they are well-defined physical observables in this effective theory approach to quantum gravity. We also find that parity-odd corrections can turn on the multipole moments that vanish in Einstein theory, which may have significant observational implications.

\vfill {\footnotesize maliang0@tju.edu.cn\ \ \ pangyi1@tju.edu.cn\ \ \ mrhonglu@gmail.com}

%{\footnotesize \hoch{*}Corresponding author}

\thispagestyle{empty}
\pagebreak

\tableofcontents
\addtocontents{toc}{\protect\setcounter{tocdepth}{2}}

%\newpage

\section{Introduction}

A century after its discovery, General Relativity has proven remarkably successful in explaining phenomena across a vast range, from the solar system to the entire universe. On the other hand, it is also widely believed that Einstein gravity should only appear as the leading term in an effective theory of quantum gravity below certain ultra-violate cutoff scale $\Lambda_c$.  In the simplest scenario, the effective action is given by an infinite derivative expansion controlled by powers of $\Lambda_c$
\bea
S=\frac{1}{16\pi G_N}\int d^4x\sqrt{-g}\left(\mathcal{L}_{2\partial}+\Lambda_c^{-2}\mathcal{L}_{4\partial}+
\Lambda_c^{-4}\mathcal{L}_{6\partial}+\cdots
\right).
\label{effA}
\eea
The higher derivative corrections to infrared physical quantities encode the fine structure of quantum gravity effects, and their observation could shed lights on hidden dynamics at the cutoff scale.  Although the correction terms are usually rather small and may not be visible in the near future, one can still ask the interesting question: which physical quantities computed using the effective action \eqref{effA} are genuine physical observables? The answer to this question is crucial for their potential measurements by future detectors.
%%%%%%%%%%%%%%%%%%%%%%%%%%%%%%%%%%%%%%%%%%%%%%%%%%%%%%%%%%%%%%%%%%%%%%%%%%%%%%%%%%%%%%%

In an effective field theory (EFT), the equivalence theorem states that meaningful physical observables should be invariant under field redefinitions, which are invertible and retain the same physical spectrum \cite{Itzykson:1980rh}. In the perturbative higher derivative extensions of Einstein gravity, field redefinitions satisfying the equivalence theorem take the form
\be
g_{\mu\nu} \rightarrow g_{\mu\nu} + \lambda_1 R_{\mu\nu} + \lambda_2 R g_{\mu\nu} +\cdots
\label{redef1}
\ee
where ``$\cdots$" refers to matter dependent terms, as well as even higher order terms. It is obvious that coefficients of certain terms in \eqref{effA} are shifted by \eqref{redef1}. Based on the Reall-Santos method \cite{Reall:2019sah}, we have proven that the Euclidean action of asymptotically flat or AdS black holes are invariant under field redefinitions \cite{Ma:2023qqj, Hu:2023gru}. Consequently, all the thermodynamic variables derived from the Euclidean action also enjoy this property. In particular, for asymptotically flat black holes, the corrections to thermodynamic variables depend only on the coupling coefficients of the higher order operators that are inert under \eqref{redef1}.
%%%%%%%%%%%%%%%%%%%%%%%%%%%%%%%%%%%%%%%%%%%

Similar to black hole thermodynamics, black hole multipole moments will also receive higher-derivative corrections, except for the mass monopole and the current dipole moment, corresponding to total mass and angular momentum that are typically chosen to be fixed under the perturbation. Black hole multipole moments play a crucial role in describing the external fields and gravitational wave radiation, potentially offering a new window into the footprints of quantum gravity \cite{Bena:2020see, Bianchi:2020bxa, Bianchi:2020miz, Bah:2021jno} beyond standard General Relativity. Upcoming observations of gravitational waves from extreme mass-ratio inspirals (EMRIs) \cite{LISA:2017pwj,Barack:2006pq,Cardoso:2016ryw,Babak:2017tow,Ryan:1995wh,Krishnendu:2018nqa} may reveal whether black hole mass and spin multipole moments match predictions from classical general relativity or suggest modifications induced by new physics beyond the standard model. It is thus urgent to understand in an effective theory of gravity, whether black hole multipole moments are meaningful physical observables, i.e. invariant under field redefinitions. Based on traditional approach, to obtain black hole multipole moments, one needs to first solve for higher derivative corrections to rotating black holes which is technically quite difficult. Until now, there is no proof that in a generic theory of gravity with higher derivative corrections, the black hole multipole moments are invariant under field redefinitions, except for very few specific examples. For instance, the recent work \cite{Cano:2022wwo} has computed the multipole moments of Kerr black holes in pure gravity with cubic curvature corrections and showed that they are invariant under field redefinitions of the metric utilizing the Ricci flatness of Kerr solution. For non Ricci flat solutions, such as Kerr-Newman black holes, the situation is unknown.
%%%%%%%%%%%%%%%%%%%%%%%%%%%%%%%%%%%%%%%%%%

In this work, we make the first attempt to study leading higher derivative corrections to Kerr-Newman black hole, with focus on its multipole moments. In four dimensions, the leading higher-derivative extensions of Einstein-Maxwell theory consist of parity even and odd 4-derivative terms built from Riemann tensor and the $U(1)$ field strength. Different from the static black hole solutions, there is no mature method of deriving the complete form of even the leading
higher-derivative corrections to rotating black holes. We thus adopt the approximate method proposed in \cite{Cardoso:2018ptl, Cano:2019ore, Cano:2022wwo}, by expanding the perturbed metric and $U(1)$ gauge potential in power series of the dimensionless parameters including $\chi_a:=a/\mu$ and $\chi_Q:=Q/\mu$, where $\mu,\,a,\,Q$ parametrize the mass, spin and electric charge of the uncorrected black hole solution respectively.

We therefore need to be concerned with two perturbative expansions of the the Kerr-Newman black hole. One is the leading-order perturbation of the 4-derivative couplings. After deriving these linear perturbed field equations, in principle, we should be able to solve them order by order in $\chi$'s up to arbitrarily high order. However, in practice, this second procedure is rather time consuming and we have to terminate at certain order.

At the first order in 4-derivative couplings, corrections from the parity even and odd terms to the black hole solution decouple from each other. Thus we can analyze these two cases separately. When parity preserving 4-derivative terms are switched on, we are able to obtain the perturbed solution up to ${\cal O}(\chi^7)$, while concerning only parity odd 4-derivative interactions, we can reach ${\cal O}(\chi^8)$ \footnote{Here we can only reach ${\cal O}(\chi^7)$ or ${\cal O}(\chi^8)$, which seems low compared to the vacuum Kerr case \cite{Cano:2019ore}. This is due to the limited computation power and the further complexity of the field equations caused by more variables and higher derivative terms switched on. }. From the approximate solution, we could read off corrections to both the gravitational and electromagnetic multipole moments at first few levels. Similar to the pure gravity case, parity preserving 4-derivative interactions only modify multipole moments \(\{M_{2n}, \mathcal{S}_{2n+1}, \mathcal{Q}_{2n}, \mathcal{P}_{2n+1}\}\) that are already nonzero at the leading order. In the parity odd case, the 4-derivative interactions contribute to multipole moments \(\{M_{2n+1}, \mathcal{S}_{2n}, \mathcal{Q}_{2n+1}, \mathcal{P}_{2n}\}\) that vanish at the leading order. Thus, their very presence breaks the equatorial symmetry of the solution and has significant observational implications (see \cite{Fransen:2022jtw} for a recent study and references therein). Most importantly, we show that these results can be expressed in a way that is manifestly invariant under field redefinitions.

%in the basis of homogeneous binary polynomials made of
% once properly choosing the integration constants in the solution so that the mass, angular momentum and electric charge remain uncorrected.

This paper is organized as follows. In section 2, we study the electrically-charged Kerr-Newman black hole and obtain the mass and current multipole moments from the metric and the electric and magnetic multipole moments from the Maxwell field. In section 3, we consider the (leading) 4-derivative corrections, which can be categorized as parity even and parity odd terms. The leading order correction is solved order by order in terms of appropriate two dimensionless parameters $(\chi_a,\chi_Q)$ of the Kerr-Newman black hole. This allows us to read off the 4-derivative corrections to the multipole moments. We then show that the results are independent of the field redefinition, indicating that they are indeed good physical quantities in our effective theory approach to quantum gravity.  We conclude the paper in section 4. In appendix A, we show how the corrections of the solution modify the thermodynamic variables, and the results are consistent with the Reall-Santos method. In appendices B and C, we give the complicated expressions that would be a digression if given in the main text. In appendix D, we discuss briefly the Geroch-Hansen method of multipole moments for general theories of gravity.

\section{Multipole moments of Kerr-Newman black hole in 2-derivative theory}

Currently, there are three different procedures to calculate black hole gravitational multipole moments, including the Geroch-Hansen formalism \cite{Geroch:1970cd,Hansen:1974zz}, the Thorne's formalism \cite{Thorne:1980ru} and the covariant phase space approach proposed in \cite{Compere:2017wrj}. Equivalence of the results obtained from three different methods has been discussed in \cite{Compere:2017wrj, Cano:2022wwo}. Here we will adopt Thorne's formalism which directly extracts gravitational multipole moments by recasting the solution in the asymptotically Cartesian and mass-centered (ACMC) coordinate system. In this section, we briefly review how to obtain the gravitaional multipole moments of Kerr-Newman black holes by working in the ACMC coordinate system.

The electrically-charged Kerr-Newman black hole is an exact solution of the Einstein-Maxwell theory, with
\be
\mathcal{L}_{2\partial}=R-\frac{1}{4}F_{\mu\nu}F^{\mu\nu}\ .
\ee
The solution takes the form
\bea
ds^2&=&-\frac{\Delta_r}{\Sigma}\left(dt-a(1-x^2)d\varphi\right)^2+
\frac{\Sigma}{\Delta_r}dr^2+\frac{\Sigma}{1-x^2}dx^2
+\frac{1-x^2}{\Sigma}\left(adt-(r^2+a^2)d\varphi\right)^2\nn\\
%%%
A_{(1)}&=&-\frac{2Qr}{\Sigma}\left(dt-a(1-x^2)d\varphi\right)\ ,\nn\\
%%%%
\Delta_r&=&r^2-2\mu r+Q^2+a^2,\qquad \Sigma=r^2+a^2x^2\ .
\label{Kerr-Newman}
\eea
 To compute all the gravitational multipole moments, we must perform a coordinate transformation from Boyer-Lindquist coordinates \((r, x)\) to ACMC-\(\infty\) coordinates \((r_S, x_S)\) defined by \cite{Bena:2020uup,Cano:2022wwo}
\be
r_S\sqrt{1-x_S^2}=\sqrt{r^2+a^2}\sqrt{1-x^2},\qquad r_Sx_S=rx\ ,
\label{ACMC leading}
\ee
in terms of which, the metric takes the form in the far zone \cite{Thorne:1980ru}
\bea
g_{tt}&=&-1+\frac{2M}{r}+\sum_{\ell\geq1}^{\infty}\frac{2}{r^{\ell+1}}\left({M}_\ell P_\ell+\sum_{\ell'<\ell}c_{\ell\ell'}^{(tt)}P_{\ell'}\right),\nn\\
%%%
g_{t\varphi}&=&-2r(1-x^2)\left[\sum_{\ell\geq1}^{\infty}\frac{1}{r^{\ell+1}}\left(\frac{{\mathcal{S}}_\ell}{\ell}P'_\ell
+\sum_{\ell'<\ell}c_{\ell\ell'}^{(t\varphi)}P_{\ell'}'\right)
\right],\cr
%%%
g_{rr}&=&1+\sum_{\ell\geq0}^{\infty}\frac{1}{r^{\ell+1}}\sum_{\ell'\leq\ell}c_{\ell\ell'}^{(rr)}P_{\ell'},\quad
g_{xx}=\frac{r^2}{1-x^2}\left[1+\sum_{\ell\geq0}^{\infty}\frac{1}{r^{\ell+1}}\sum_{\ell'\leq\ell}c_{\ell\ell'}^{(xx)}P_{\ell'}
\right],\nn\\
%%%
g_{\varphi\varphi}&=&r^2(1-x^2)\left[1+\sum_{\ell\geq0}^{\infty}\frac{1}{r^{\ell+1}}\sum_{\ell'\leq\ell}c_{\ell\ell'}^{(\varphi\varphi)}P_{\ell'}
\right],\quad g_{rx}=r\left[\sum_{\ell\geq0}^{\infty}\frac{1}{r^{\ell+1}}\sum_{\ell'\leq\ell}c_{\ell\ell'}^{(rx)}P_{\ell'}'
\right]\ ,
\label{AC-N}
\eea
where $P_\ell$ and $P_\ell'$ denote Legendre polynomial of $x$ and its $x$-derivative, respectively. Moreover we have removed ``$S$'' from the subscript to simplify the notation. The coefficients $M_{\ell}$ and $\mathcal{S}_{\ell}$ are the mass and current multipole moments respectively. On the other hand, the coefficients $c^{(ij)}_{\ell\ell'}$ are gauge dependent and nonphysical \cite{Thorne:1980ru}. For the Kerr-Newman black hole, the nonvanishing multiplet moments are
%%%%
\be
M_{2n}=\mu(-a^2)^n,\qquad \mathcal{S}_{2n+1}=\mu a(-a^2)^n\ ,
\label{multipole 2}
\ee
%%%
where $M_0$ and $\mathcal{S}_1$ correspond to the total mass and angular momentum.
It has been noted in \cite{Bena:2020uup,Sotiriou:2004ud} that the multipole moments of the Kerr-Newman black hole share the same form as those of the Kerr black hole in pure Einstein gravity. In other words, they are unaffected by the electric charge.

In ACMC-\(\infty\) coordinates \((r_S, x_S)\), we find that the Maxwell field in the far zone takes the form, after dropping the subscript ``$S$'',
\bea
A_t&=&-\sum_{\ell\geq0}^{\infty}\frac{4}{r^{\ell+1}}\Big({\mathcal{Q}}_\ell P_\ell+\sum_{\ell'<\ell}c_{\ell\ell'}^{(t)}P_{\ell'}\Big)\ ,
\nn\\
%%%
A_\varphi&=&\sum_{\ell\geq0}^{\infty}\Bigg\{\frac{4x}{r^{2\ell}}\Big({\mathcal{P}}_{2\ell} P_{2\ell}+\sum_{\ell'<\ell}c_{2\ell,2\ell'}^{(\varphi)}P_{2\ell'}\Big)\cr
&&-\frac{1-x^2}{r^{2\ell+1}}\frac{4}{2\ell+1}\Big({\mathcal{P}}_{2\ell+1} P_{2\ell+1}'+\sum_{\ell'<\ell}c_{2\ell+1,2\ell'+1}^{(\varphi)}P_{2\ell'+1}'\Big)
\Bigg\}\ ,
\label{ACMC ele}
\eea
from which we recognize $\mathcal{Q}_\ell$ as the electric multipole moments are given by
\be
\mathcal{Q}_{2n}=\frac{1}{2}Q(-a^2)^n,\qquad {\cal Q}_{2n+1}=0\ .
\label{ele multipole 2}
\ee
In particular, the electric charge is given by $Q_e=\mathcal{Q}_0 =Q/2$. The coefficients ${\cal P}_{\ell}$ take the values below
\be
{\cal P}_{2n+1}=-\frac{1}{2}Qa(-a^2)^n,\qquad {\cal P}_{2n}=0\ .
\ee
As we shall show below that ${\cal P}_{\ell}$ can be interpreted as the magnetic multipole moments, because  via electromagnetic duality, they appear in the electric multipole moments of the dual $U(1)$ gauge field. Moreover, in the dual $U(1)$ gauge field, the coefficients in front of $xP_{2n}$ are in fact non-zero and given by $\mathcal{Q}_{2n}$. This is why we introduce the first line in the expansion of $A_{\varphi}$ to make the ansatz more general, even though for the solution given in \eqref{Kerr-Newman}, it is absent in the expansion of $A_{\varphi}$. Note that in this paper, we consider only the electrically charged Kerr-Newman black hole, the magnetic monopole ${\cal P}_0$ or the magnetic charge vanishes. The electric charge, under the rotation, can generate odd magnetic multipole moments, namely ${\cal P}_{2n+1}$.

To ensure the gauge invariance of the electric multiple moments obtained above, we define electric potential rigorously using the Killing vector $\xi=\partial_t+\Omega_H\partial_\varphi$ that vanishes on the bifurcation horizon of the black hole.
\be
\xi^{\mu}F_{\mu\nu}=\partial_{\nu}\Phi_e\quad\Longrightarrow\quad \Phi_e=A_t+\O_H A_{\varphi}+{\rm const.}
\ee
Thus in the far-zone, the large $r$ expansion of the gauge invariant electric potential $\Phi_e$ acquires not only terms proportional to Legendre polynomials as in the static case, but also terms proportional to derivatives of Legendre polynomials, due to frame dragging effects.

We now examine the magnetic multipole moments more closely. To define them in a rigorous way, we consider the dual $U(1)$ gauge field $\widetilde{A}_{\mu}$ whose field strength is the Hodge dual of $F_{\mu\nu}$, i.e.,
$\widetilde{F}_{\mu\nu}=\ft12\epsilon_{\mu\nu\rho\lambda}F^{\rho\lambda}$. Therefore we have
%%%%%
\bea
\widetilde{A}_{(1)}&=&-2\frac{Qax}{\Sigma}dt+
2\Big(\frac{Qa^2(1-x^2)}{\Sigma}+Q\Big)xd\varphi\ ,
\eea
whose components take the form in the ACMC coordinate system
%%%%%
\bea
\widetilde{A}_t&=&-\sum_{\ell\geq0}^{\infty}\frac{4}{r^{\ell+1}}\Big({\widetilde{\mathcal{Q}}}_\ell P_\ell+\sum_{\ell'<\ell}\widetilde{c}_{\ell\ell'}^{(t)}P_{\ell'}\Big)\ ,
\nn\\
%%%
\widetilde{A}_\varphi&=&\sum_{\ell\geq0}^{\infty}\Bigg\{\frac{4x}{r^{2\ell}}\Big({\widetilde{\mathcal{P}}}_{2\ell} P_{2\ell}+\sum_{\ell'<\ell}\widetilde{c}_{2\ell,2\ell'}^{(\varphi)}P_{2\ell'}\Big)\cr
&&-\frac{1-x^2}{r^{2\ell+1}}\frac{4}{2\ell+1}\Big({\widetilde{\mathcal{P}}}_{2\ell+1} P_{2\ell+1}'+\sum_{\ell'<\ell}\widetilde{c}_{2\ell+1,2\ell'+1}^{(\varphi)}P_{2\ell'+1}'\Big)
\Bigg\}\ ,
\label{ACMC mag}
\eea
with
\bea
{\widetilde{\cal Q}}_{2n+1}=-{\cal P}_{2n+1},\quad {\widetilde{\cal Q}}_{2n}=-{\cal P}_{2n}=0,\quad
\widetilde{\mathcal{P}}_{2n}={\cal Q}_{2n},\quad {\widetilde{\cal P}}_{2n+1}={\cal Q}_{2n+1}=0\ .
\eea
Thus we see that the coefficients ${\cal P}_{\ell}$ that appear in the expansion of $A_{\mu}$ indeed correspond to the magnetic multipole moments. Similarly, we can also define gauge invariant magnetic potential as
\be
\xi^{\mu}\widetilde{F}_{\mu\nu}=\partial_{\nu}\Phi_m\quad\Longrightarrow \quad \Phi_m=\xi^\mu \widetilde{A}_\mu+{\rm const.}\ ,
\ee
which justifies the gauge invariance of the expansion coefficients in \eqref{ACMC mag}.

%%%%%%%%%%%%%%%%%%%%%%%%%%%%%%%%%%%%%%%%%%%%%%%%%%%%%%%%%%%%%%%%%

\section{Adding 4-derivative corrections }

In this section, we will apply the strategy of \cite{Cano:2022wwo} to compute generic 4-derivative corrections to the multipole moments of $D=4$ Kerr-Newman black hole. The  4-derivative extension of Einstein-Maxwell theory contains parity even and odd terms. Up to first order in 4-derivative couplings, their corrections to the black hole solution and multipole moments are disentangled from each other. Thus we shall discuss the parity even and odd cases separately.

\subsection{Parity-even case}

The general parity-even 4-derivative interactions involving curvature and Maxwell field strength are given by
\bea
 \mathcal{L}^{(e)}_4 &=& c_1 R^2 + c_2 R^{\mu\nu} R_{\mu\nu} + c_3 R^{\mu\nu\rho\sigma} R_{\mu\nu\rho\sigma} + c_4 R F^2 + c_5 R^{\mu\nu} F_{\mu\rho} F_\nu{}^\rho\nn\\
&&+ c_6 R^{\mu\nu\rho\sigma} F_{\mu\nu} F_{\rho\sigma} + c_7 (F^2)^2 + c_8 F^{\mu}{}_\nu F^\nu{}_\rho F^\rho{}_\sigma F^{\sigma}{}_\mu \,.
\label{even}
\eea
It is interesting to note that the following structures are independently invariant under the electromagnetic duality transformation \cite{Charles:2017dbr}
%%%
\bea
&& F^{\mu}{}_\nu F^\nu{}_\rho F^\rho{}_\sigma F^{\sigma}{}_\mu-\frac14 (F^2)^2,\quad R^{\mu\nu} F_{\mu\rho} F_\nu{}^\rho-\frac14 RF^2\ ,
\nn\\
&& R_{\m\n\r\s}R^{\m\n\r\s}-4R_{\m\n}R^{\m\n}+R^2,\quad R_{\m\n}R^{\m\n},\quad R^2\ .
\eea

%%%
The most general redefinition of the metric that preserves the parity is of the form
\bea
g_{\mu\nu} \rightarrow g_{\mu\nu} + \lambda_1 R_{\mu\nu} + \lambda_2 R g_{\mu\nu} +
\lambda_3 F_{\mu\rho} F_{\nu}{}^\rho + \lambda_4 F^2 g_{\mu\nu}\,,\label{redef}
\eea
which leads to the variation of 4-derivative coefficients as
\bea
&&c_{1}  \rightarrow  c_{1}+\ft12\lambda_{1}+  \lambda_{2}\,,\qquad
c_{2}  \rightarrow  c_{2}-\lambda_{1}\,,\qquad c_{3}  \rightarrow  c_{3}\,,\nn\\
&&c_{4}\rightarrow  c_{4}-\ft18\lambda_{1}+\ft12\lambda_{3}+
\lambda_{4}\,,\quad
c_{5} \rightarrow c_{5}+\ft12\lambda_{1}-\lambda_{3}\,,\qquad c_{6}  \rightarrow c_{6}\,,\nn\\
&&c_{7}  \rightarrow c_{7}-\ft18\lambda_{3}\,,\qquad
c_{8}   \rightarrow c_{8}+\ft12\lambda_{3}\ ,
\eea
under which the combinations below
\bea
\alpha_0&=&2 c_2+8 c_3+4 c_5+4 c_6+32 c_7+16 c_8\ ,
\nn\\
\alpha_1&=&c_3,\qquad \alpha_2=c_6,\qquad \alpha_3=c_2+2 c_5+4 c_8\ ,\label{alphas}
\eea
are invariant \cite{Cheung:2018cwt,Ma:2020xwi}. Thus a physical quantity depending only on these combinations satisfy our criteria of being a meaningful observable.

For the time being, there is no well established approach to
finding  analytical and complete results for 4-derivative corrections to a rotating black hole. Hence we resort to the approximate method proposed by \cite{Cano:2019ore}. Up to first order in $c_i$, we recast the corrected field equations in the form \cite{Hu:2023qhs}
%%%%%%%
\bea
R_{\m\n}-\ft12 g_{\m\n}R&=&\ft12 (F_{\mu}^{\ \rho}F_{\nu\rho}-\frac{1}{4}g_{\mu\nu}F^2)+\ft12 \Delta T_{\m\n}[g^{(0)}_{\l\s}\,,A^{(0)}_\s]\ ,
\nn\\
\nabla_{\mu}F^{\mu\nu}&=&\Delta J^\nu[g^{(0)}_{\l\s}\,, A^{(0)}_\s]\ ,
\label{4deom}
\eea
where $g^{(0)}_{\m\n}$ and $A^{(0)}_\m$ denote the uncorrected solution and the effective energy-momentum tensor and the effective electric current are defined by
\be
\Delta T_{\m\n}=-\frac{2}{\sqrt{-g}}\frac{\delta{(\sqrt{-g}\cal L}^{(e)}_4)}{\delta g^{\mu\nu}}\,,\qquad \Delta J^\nu=-\frac{1}{\sqrt{-g}}\frac{\delta{(\sqrt{-g}\cal L}^{(e)}_4)}{\delta A_{\nu}}\ .\label{effective tensor}
\ee
%%%%%%%
Next, when solving for the perturbed solution, we expand the uncorrected solution in power series of $\chi_a = a / \mu,\,\chi_Q=Q/\m$ both smaller than 1. Depending on the available computer power, one can solve for the perturbed solution up to certain order in $\chi$. Since the leading order solution is exact in $\chi$ and the approximation is only performed in finding the perturbed solution,  it has been shown that \cite{Cano:2019ore, Cano:2022wwo} this procedure can yield a rather accurate approximation to the full perturbed solution, as long as it is carried out to sufficiently high order in $\chi$.

Since the uncorrected Kerr-Newman black hole is stationary and axisymmetric, the effective energy-momentum tensor and electric current will inherit these symmetries. Consequently, the perturbed black hole solution can be parameterized as \cite{Cano:2019ore}
%%%%%%
\bea
ds^2&=&-\frac{\Delta_r}{\Sigma}\left(dt-a(1-x^2)d\varphi\right)^2(1+H_1)+(1+H_2)\left(\frac{\Sigma}{\Delta_r}dr^2+\frac{\Sigma}{1-x^2}dx^2\right)\cr
&&+\frac{1-x^2}{\Sigma}\left((1+H_3)adt-(1+H_4)(r^2+a^2)d\varphi\right)^2\cr
%%%
A_{(1)}&=&-\frac{2Qr}{\Sigma}\left((1+H_5)dt-(1+H_6)a(1-x^2)d\varphi\right)\ ,
\label{perturbed solution}
\eea
%%%%
where the six functions $H_{i}$, $i=1,\dots 6,$ depend only on coordinates $r$ and $x$. At first glance, the form of the metric ansatz appears different from the original one used in \cite{Cano:2019ore}, but one can easily show that the two ans\"{a}tze are related via
\bea
H_1^{\rm CR}&=&\frac{2a^2 H_3 (1-x^2)-H_1 \Delta _r}{\Sigma },\quad
H_2^{\rm CR}=\frac{(H_3+H_4) (r^2+a^2)-H_1 \Delta _r}{2 \mu  r}\ ,
\nn\\
%%%
H_3^{\rm CR}&=& H_2,\quad %%%
H_4^{\rm CR}=-\frac{a^2 H_1 \Delta _r(1-x^2) -2H_4  (r^2+a^2)^2}{\Sigma  (r^2+a^2)+2 a^2 \mu  r (1-x^2)}\ ,
\eea
where the superscript ``${\rm CR}$" denotes the $H_i$'s defined in \cite{Cano:2019ore}. In the next, we shall expand each $H_i$ in power series of parameters
%%%%
\be
\chi_a=\frac{a}{\mu},\qquad \chi_Q=\frac{Q}{\mu}\ ,\quad \hbox{with}\quad\chi_a^2 + \chi_Q^2 \le 1\,,
\ee
%%%%
where the right inequality is saturated by the extremal unperturbed Kerr-Newman black hole. In an infinite series expansion, there are many ways to recollect terms. For simplicity, we choose homogeneous polynomials of $\chi_a$ and $\chi_Q$ as the expansion basis. To do so, we introduce a bookkeeping parameter $\epsilon$ which will be set to 1 at the end of calculation and temporarily replace $\chi$ by $\epsilon\chi$. Then the expansion of $H_i$ can be conveniently written as
%%%%%%%%%
\be
H_i(r,x)=\sum_{n=0}^\infty H_i^{(n)}(r,x)\epsilon^n\ .
\label{Hi}
\ee
%%%%%%%%%
Following \cite{Cano:2019ore}, $H_i^{(n)}(r,x)$ can
always be expressed as a polynomial in $x$ and in $1/r$
\be
H_i^{(n)}(r,x)=\sum_{p=0}^n\sum_{k=0}^{k_{\rm max}}H_i^{(n,p,k)}\frac{x^p}{r^k}\ ,
\label{Hin}
\ee
where $H_i^{(n,p,k)}$ are constant coefficients and for each undetermined function, the number of $k_{\rm max}$ depends on $n$ and $p$. Substituting \eqref{perturbed solution}, \eqref{Hi} and \eqref{Hin} into \eqref{4deom}, we solve for $H_i^{(n,p,k)}$ order by order in $\epsilon$ up to ${\cal O}(\epsilon^7)$ and present the results in an accompanying Mathematica notebook. As a double check of our approximate solution, in Appendix \ref{thermo}, we first calculate 4-derivative corrections to all the thermodynamic variables of Kerr-Newman black hole using the Reall-Santos method \cite{Reall:2019sah}, which requires only the knowledge of the uncorrected solution. We then compute the same quantities by applying the standard approach to the corrected solution. It turns out that  results obtained from these two methods agree with each other up to the approximation order we have considered.

The higher-derivative terms also modifies the relation between the ACMC-$\infty$ coordinate denoted by $(r_S,\, x_S)$ and the Boyer-Lindquist coordinates \((r, x)\)
%%%%%
\bea
r&=&r^{(0)}+ r^{(1)}(r_S,x_S),\quad x=x^{(0)}+ x^{(1)}(r_S,x_S)\ ,
\eea
%%%%
where the leading terms $r^{(0)}$ and $x^{(0)}$ are solved from \eqref{ACMC leading} yielding
%%%%%
\bea
r^{(0)}&=&\frac{\sqrt{r_S^2-a^2+\sqrt{a^4+2 a^2 r_S^2 \left(2 x_S^2-1\right)+r_S^4}}}{\sqrt{2}}\ ,
\nn\\
x^{(0)}&=&\frac{\sqrt{2} r_S x_S}{\sqrt{r_S^2-a^2+\sqrt{a^4+2 a^2 r_S^2 \left(2 x_S^2-1\right)+r_S^4}}}\ ,
\eea
%%%%%
while the corrections caused by 4-derivative interactions takes the form
\bea
 r^{(1)}(r_S,x_S)=\sum_{k=-1}^{\infty}\sum_{p=0}^{k+1}b_{k,p}\frac{x_S^p}{r_S^k}\ ,\qquad
 %%%
 x^{(1)}(r_S,x_S)=(1-x_S^2)\sum_{k=1}^{\infty}\sum_{p=0}^{k-1}c_{k,p}
 \frac{x_S^p}{r_S^k}\ .
\eea
The coefficients $b_{k,p}$ and $c_{k,p}$ are fixed by requiring the modified metric to remain in the standard form \eqref{AC-N}, from which we read off corrections to
the gravitational multipole moments due to the parity even 4-derivative interactions. In terms of new integration constants $\m',\,\chi_a',\,\chi_Q'$ defined below
%%%%%%%%%
\bea
\mu\rightarrow\mu'=\mu+\delta\mu,\quad \chi_a\rightarrow\chi_a'=\chi_a+\delta \chi_a,\quad \chi_Q\rightarrow\chi_Q'=\chi_Q+\delta \chi_Q\ ,
\label{parameter redefinition}
\eea
where $\d\m,\,\d\chi_a,\,\d\chi_Q$ are given in Appendix \ref{redefin}, we
find that the mass, angular momentum, and electric charge of the Kerr-Newman black hole become independent of 4-derivative couplings. Meanwhile, all the multipole moments are manifestly invariant under field redefinitions. We also note that corrections to mass multipole moments behave as
$\delta M_{n} \sim \chi^{n+2}$ (+higher order in $\chi$). Therefore, in the parity-even case, with the perturbed solution up to ${\cal O}(\chi^7)$, we can only obtain non-trivial corrections up to $M_4$.

Below we present the 4-derivative corrections to multipole moments written in terms of the new integration constants, and for the tidiness, we remove the ``prime" from notation. Using the approximate solution, we are able to obtain the first few gravitational multipole moments listed below
%%%%%%%%%%%%%%%%%%%%%
\bea
\delta M_2^{(e)}&=&\alpha _2 \mu  \chi _a^2 \chi _Q^2-\frac{\mu   \chi _a^2 \chi _Q^2}{300}  \Big[(8 \alpha_0-76 \alpha _1-203 \alpha_2-19 \alpha _3)  \chi _Q^2+30 \alpha_2 \chi _a^2\Big]\ ,
\nn\\
\delta M_4^{(e)}&=&-\frac{2402  \mu ^3  \chi _a^4 \chi _Q^2}{1225}\alpha_2\ ,
\qquad \delta M_6^{(e)}=0\ ,
\nn\\
\delta \mathcal{S}_3^{(e)}&=&\frac{23}{25} \alpha _2 \mu ^2 \chi _a^3 \chi _Q^2-\frac{\mu ^2  \chi _a^3 \chi _Q^2 }{4900}\Big[\alpha _2 (637 \chi _a^2-3501 \chi _Q^2)+(102 \alpha _0-1172 \alpha _1-293 \alpha _3) \chi _Q^2\Big]\ ,
\nn\\
\delta \mathcal{S}_5^{(e)}&=&-\frac{453}{245} \alpha_2 \mu ^4  \chi _a^5 \chi _Q^2\ ,\qquad \delta \mathcal{S}_7^{(e)}=0\ .
\eea
For electric and magnetic multipole moments, we have
\bea
\delta \mathcal{Q}_2^{(e)}&=&\frac{3}{50} \alpha _2 \mu  \chi _a^2 \chi _Q+\frac{\mu  \chi _a^2 \chi _Q}{4200}\Big[\alpha _2 (90 \chi _a^2+1519 \chi _Q^2)-49 (\alpha _0+4 \alpha _1+\alpha _3) \chi _Q^2\Big]\nn\\
&&+\frac{\mu  \chi _a^2 \chi _Q}{94080}\Big[
980 \alpha _2 \chi _a^4-3 (197 \alpha _0+1016 \alpha _1+3288 \alpha _2+254 \alpha _3) \chi _a^2 \chi _Q^2\nn\\
&&-392 (4 \alpha _0-20 \alpha _1-64 \alpha _2-4\alpha _3) \chi _Q^4
\Big]\ ,
\nn\\
\delta \mathcal{Q}_4^{(e)}&=&-\frac{97 \alpha _2 \mu ^3 \chi _a^4 \chi _Q}{1225}+\frac{\mu ^3 \chi _a^4 \chi _Q}{411600}\Big[
(6905 \alpha _0+29256 \alpha _1-320788 \alpha _2+7314 \alpha _3) \chi _Q^2\nn\\
&&-10976 \alpha _2 \chi _a^2
\Big]\ ,
\nn\\
\delta \mathcal{Q}_6^{(e)}&=&\frac{125 \alpha _2 \mu ^5 \chi _a^6 \chi _Q}{1386}\ ,
\nn\\
%%%
\delta \mathcal{P}_1^{(e)}&=&-\frac{1}{2} \alpha _2 \chi _a \chi _Q+\frac{\chi _a \chi _Q}{120}\Big[
\alpha _2 (6 \chi _a^2-13 \chi _Q^2)+(\alpha _0-20 \alpha _1-5 \alpha _3) \chi _Q^2\Big]
+\frac{\chi _a \chi _Q}{3360}\Big[
63 \alpha _2 \chi _a^4
\nn\\
&&-6 (4 \alpha _0+4 \alpha _1+27 \alpha _2+\alpha _3) \chi _a^2 \chi _Q^2+14 (\alpha _0-20 \alpha _1-10 \alpha _2-5 \alpha _3) \chi _Q^4\Big]\ ,
\nn\\
%%%
\delta \mathcal{P}_3^{(e)}&=&\frac{13}{25} \alpha _2 \mu ^2 \chi _a^3 \chi _Q-\frac{\mu ^2 \chi _a^3 \chi _Q}{2450}\Big[
112 \alpha _2 \chi _a^2+(11 \alpha _0-528 \alpha _1+755 \alpha _2-132 \alpha _3) \chi _Q^2\Big]\ ,
\nn\\
%%%
\delta \mathcal{P}_5^{(e)}&=&-
\frac{781 \alpha _2 \mu ^4 \chi _a^5 \chi _Q}{1470}\ .\label{ele multipole even}
\eea
The results depends only on the invariant combinations $(\alpha_0,\alpha_1,\alpha_2, \alpha_3)$ of coupling constants defined in \eqref{alphas}. It is evident that the parity even 4-derivative terms will not contribute nontrivially to $M_{2n+1}$ and ${\cal S}_{2n}$.

Similar to the 2-derivative case, from the field equation of $A_\m$, we can define its magnetic dual vector field $\widetilde{A}_\m$
%%%%
\be
\nabla^{\m}{\cal D}^{(e)}_{\m\n}=0\quad \Longrightarrow\quad
\frac12\epsilon_{\m\n\r\l} {\cal D}^{(e)\r\l}=2\nabla_{[\m}\widetilde{A}^{(e)}_{\n]}\ .
\ee
%%%%
The explicit form of the induction tensor $D^{(e)}_{\m\n}$ can be seen in Appendix \ref{eom} from the parity even part of the total $U(1)$ field equation. We find that
%%%%
\be
\widetilde{A}_{(1)}^{(e)}=-2\Big(\frac{Qax}{\Sigma}+\widetilde{\mathcal{H}}_{1}\Big)dt-2\Big(-\frac{Qa^2x(1-x^2)}{\Sigma}-Qx+\widetilde{\mathcal{H}}_{2}\Big)d\varphi\ ,
\ee
%%%%
where $\mathcal{H}_{1}$ and $\mathcal{H}_{2}$ are also expanded in terms of power series of $\chi_a$ and $\chi_Q$. Readers interested in their explicit forms up to ${\cal O}(\chi^7)$ are referred to the accompanying Mathematica file. Matching $\widetilde{A}_{(1)}$ to the desired form in \eqref{ACMC mag}, we read off the leading expansion coefficients which confirm the relation

%%%%%
\be
\delta \widetilde{\mathcal{P}}_{0,\,2,\,4,\,6}^{(e)}=\delta \mathcal{Q}_{0,\,2,\,4,\,6}^{(e)},\qquad \delta \widetilde{\mathcal{Q}}_{1,\,3,\,5}^{(e)}=-\delta \mathcal{P}_{1,\,3,\,5}^{(e)}\ .
\ee
%%%%%%%%

\subsection{Parity-odd case}

We now turn to study the effects of parity-odd 4-derivative interactions on Kerr-Newman black hole and its multipole moments. In $D=4$,  the independent parity odd 4-derivative terms can be parametrized as
\be
\mathcal{L}_{4}^{(o)}=d_1RF^{\mu\rho}\widetilde{F}_{\mu\rho}+d_2R_{\mu\nu\rho\sigma}F^{\mu\nu}\widetilde{F}^{\rho\sigma}+\frac{d_3}4\widetilde{F}_{\mu \nu}F^{\mu\nu}F^2\ ,
\label{odd}
\ee
where $\widetilde{F}_{\mu \nu}:=\frac{1}{2}\epsilon_{\mu\nu\rho\sigma}F^{\rho\sigma}$ and we have used the useful identity that $g_{\mu\nu}F_{\rho\sigma}\widetilde{F}^{\rho\sigma}=4F_{\mu\rho}\widetilde{F}_\nu^{\ \rho}$. Note that at 4-derivative level, there exists no parity-odd term preserving electromagnetic duality invariance. Because of the same identity, the parity odd field redefinition of the metric
has only one structure
\be
g_{\mu\nu} \rightarrow g_{\mu\nu}+
\lambda_5 g_{\m\n }F_{\rho\sigma}\widetilde{F}^{\rho\sigma} \,,
\ee
which shifts the coupling constants in \eqref{odd}
as
\be
d_1\rightarrow d_1+\lambda_5,\quad d_2\rightarrow d_2,\quad d_3\rightarrow d_3\ .
\label{oddred}
\ee
It is evident that 4-derivative couplings below are invariant  under the field redefinition
\be
\beta_0=d_3,\quad \beta_1=d_2\ .
%,\quad \beta_3=2d_4-d_1
\ee
%%%%

Repeating the same procedure as in the parity even case, we solve the perturbed solution up to ${\cal O}(\chi^8)$. The modifications to the field equations due to parity odd 4-derivative terms are given in Appendix \ref{eom}. Switching to the ACMC coordinate system, we obtain the parity odd 4-derivative contributions to the first few multipole moments. Interestingly, the $M_{2n}$, ${\cal S}_{2n+1}$, ${\cal Q}_{2n}$ and ${\cal P}_{2n+1}$ which are nonvanishing at 2-derivative level, do not receive corrections. Instead, non-trivial corrections appear in $M_{2n+1}$, ${\cal S}_{2n}$, ${\cal Q}_{2n+1}$ and ${\cal P}_{2n}$. This is understandable because the field equations are parity odd, therefore shifting the multipolar index $\ell$ by 1. The mass dipole remains zero because of the choice of ACMC coordinates. Specifically, for the gravitational multipole moments we have

%%%%%%%%%%%
\bea
\delta M_3^{(o)}&=&-\frac{23}{25} \beta _1 \mu ^2 \chi _a^3 \chi _Q^2+\frac{\mu ^2 \chi _a^3 \chi _Q^2}{700}  (91 \beta _1 \chi _a^2+142 \beta _0 \chi _Q^2-391 \beta _1 \chi _Q^2)\ ,
\nn\w1
%%%
\delta M_5^{(o)}&=&\frac{453}{245} \beta _1 \mu ^4 \chi _a^5 \chi _Q^2\ ,
\cr
%%%
\delta \mathcal{S}_{2}^{(o)}&=&\beta _1 \mu  \chi _a^2 \chi _Q^2-\frac{\mu  \chi _a^2 \chi _Q^2}{300}  (30 \beta _1 \chi _a^2+70 \beta _0 \chi _Q^2-159 \beta _1 \chi _Q^2)\nn\\
&&-\frac{\mu  \chi _a^2 \chi _Q^2}{560}  \Big[28 \beta _0 \chi _Q^2 (5 \chi _Q^2-\chi _a^2)+3 \beta _1 (5 \chi _a^2 \chi _Q^2+7 \chi _a^4-56 \chi _Q^4)\Big]\ ,
\nn\w1
%%%
\delta \mathcal{S}_{4}^{(o)}&=&-\frac{2402 \beta _1 \mu ^3 \chi _a^4 \chi _Q^2}{1225}+\frac{\mu ^3 \chi _a^4 \chi _Q^2 }{7350}(1603 \beta _1 \chi _a^2+3331 \beta _0 \chi _Q^2-7845 \beta _1 \chi _Q^2)\ ,
\nn\\
%%%
\delta \mathcal{S}_{6}^{(o)}&=&\frac{70667 \beta _1 \mu ^5 \chi _a^6 \chi _Q^2}{24255}\ ,
\eea
%%%
and for electric and magnetic multipole moments, we obtain
\bea
\delta \mathcal{Q}_{1}^{(o)}&=&-\frac{1}{2} \beta _1 \chi _a \chi _Q+\frac{\chi _a \chi _Q}{120}\Big[3 \beta _1 (2 \chi _a^2+\chi _Q^2)+14 \beta _0 \chi _Q^2\Big]
+\frac{\chi _a \chi _Q}{480}\Big[4 \beta _0 \chi _Q^2 (7 \chi _Q^2-3 \chi _a^2)\nn\\
&&+3 \beta _1 (3 \chi _a^4-2 \chi _a^2 \chi _Q^2+4 \chi _Q^4)\Big]
+\frac{\chi _a \chi _Q}{640}\Big[\beta _1 (9 \chi _a^4 \chi _Q^2-12 \chi _a^2 \chi _Q^4+6 \chi _a^6+10 \chi _Q^6)\nn\\
&&+2\beta _0 (10 \chi _Q^6-3 \chi _a^4 \chi _Q^2)\Big]\ ,
\nn\\
%%%
\delta \mathcal{Q}_{3}^{(o)}&=&\frac{13}{25} \beta _1 \mu ^2 \chi _a^3 \chi _Q-\frac{\mu ^2 \chi _a^3 \chi _Q}{350} \Big[16 \beta _1 \chi _a^2+(44 \beta _0+177 \beta _1) \chi _Q^2\Big]\nn\\
&&-\frac{\mu ^2 \chi _a^3 \chi _Q }{50400}\Big[882 \beta _1 \chi _a^4-(1142 \beta _0+3807 \beta _1) \chi _a^2 \chi _Q^2-36 (70 \beta _0-431 \beta _1) \chi _Q^4\Big]\ ,
\nn\\
%%%
\delta \mathcal{Q}_{5}^{(o)}&=&-\frac{781 \beta _1 \mu ^4 \chi _a^5 \chi _Q}{1470}
+\frac{\mu ^4 \chi _a^5 \chi _Q }{52920}\Big[2310 \beta _1 \chi _a^2+(6934 \beta _0+52011 \beta _1) \chi _Q^2\Big]\ ,
\nn\\
%%%
\delta \mathcal{Q}_{7}^{(o)}&=&\frac{121388 \beta _1 \mu ^6 \chi _a^7 \chi _Q}{225225}\ ,
\nn\\
%%%
\delta \mathcal{P}_{2}^{(o)}&=&-\frac{3}{50} \beta _1 \mu  \chi _a^2 \chi _Q-\frac{\mu  \chi _a^2 \chi _Q}{4200}\Big[
90 \beta _1 \chi _a^2-49 (2 \beta _0-39 \beta _1) \chi _Q^2\Big]\nn\\
&&-\frac{\mu  \chi _a^2 \chi _Q}{3360}\Big[35 \beta _1 \chi _a^4-3 (10 \beta _0+53 \beta _1) \chi _a^2 \chi _Q^2-28 (13 \beta _0-30 \beta _1) \chi _Q^4\Big]\ ,
\nn\\
%%%
\delta \mathcal{P}_{4}^{(o)}&=&\frac{97 \beta _1 \mu ^3 \chi _a^4 \chi _Q}{1225}+\frac{2 \mu ^3 \chi _a^4 \chi _Q}{3675}\Big[49 \beta _1 \chi _a^2-2 (29 \beta _0-843 \beta _1) \chi _Q^2\Big]\ ,
\nn\\
%%%
\delta \mathcal{P}_{6}^{(o)}&=&-\frac{125 \beta _1 \mu ^5 \chi _a^6 \chi _Q}{1386}.
\eea
It is evident that all the expressions above are manifestly invariant under field redefinitions. Note that the magnetic charge ${\cal P}_0$, which is zero in our original Kerr-Newman black hole, remains uncorrected, whilst the magnetic higher even multipole moments all receive corrections by the party-odd 4-derivative terms. Clearly, observation of the multipole moments above would indicate the presence of parity-odd terms and the breaking of electromagnetic duality invariance. 

Similar to the parity-even case, using the $U(1)$ field equations, we can also define the dual 1-form potential $\widetilde{A}^{(o)}_1$ which takes the form
%%%%%%%
\be
\widetilde{A}_{(1)}^{(o)}=-2\Big(\frac{Qax}{\Sigma}+\widetilde{\mathcal{H}}_{3}\Big)dt-2\Big(-\frac{Qa^2x(1-x^2)}{\Sigma}-Qx+\widetilde{\mathcal{H}}_{4}\Big)d\varphi\ .
\label{magA}
\ee
%%%%%%%
where $\mathcal{H}_{3}$ and $\mathcal{H}_{4}$ are also expanded in terms of power series of $\chi_a$ and $\chi_Q$. Readers interested in their explicit forms up to ${\cal O}(\chi^8)$ are referred to the accompanying Mathematica file.
In terms of the ACMC coordinates,  by matching the large $r$ expansion of \eqref{magA} to \eqref{ACMC mag}, we confirm that the electric and magnetic multipole moments satisfy the electromagnetic duality relation
%%%%%%%%%%%%%%
\be
\widetilde{\mathcal{Q}}_{1,\,3,\,5,\,7}^{(o)}=-\mathcal{P}_{1,\,3,\,5,\,7}^{(o)}\,,
\quad \delta\widetilde{\mathcal{Q}}_{2,\,4,\,6}^{(o)}=-\delta\mathcal{P}_{2,\,4,\,6}^{(o)}\ .
\ee

\section{Conclusion}

Black hole multipole moments are potentially useful observables for probing the nonlinear structures of General Relativity and its modifications due to unknown UV physics. In this work, we made progress in addressing a key question: are higher derivative perturbative corrections to black hole multipole moments invariant under field redefinitions. So far, traditional methods for calculating multipole moments have not explicitly demonstrated this property for general effective theory of gravity.  As far as we are aware, the field redefinition invariance of black hole multipole moments has only been shown for $D=4$ Einstein gravity extended by cubic curvature terms \cite{Cano:2022wwo}, where the Ricci flatness of the leading order solution had played a role in the proof. Here, we further investigated the leading higher-derivative corrections (\ref{even},\ref{odd}) to the non-Ricci flat Kerr-Newman black hole \eqref{Kerr-Newman}. We find that the leading higher derivative corrections to black hole multipole moments indeed depend only on the combinations of coupling constants that are inert under field redefinitions.

To achieve this, we first cast the electrically-charged Kerr-Newman black hole in the ACMC coordinates. We read off the nonvanishing mass and current multipole moments \(\{M_{2n}, \mathcal{S}_{2n+1}\}\) from the metric, and the nonvanishing electric and magnetic multipole moments \(\{\mathcal{Q}_{2n}, \mathcal{P}_{2n+1}\}\) from the Maxwell field. The absence of the magnetic monopole has a consequence that all the even magnetic multiple moments vanish, and the odd ones are generated by the electric charge under rotation. We then
adopted an approximate method of solving for the higher derivative corrected metric and $U(1)$ gauge field order by order in dimensionless rotation parameter $\chi_a={a}/{\m}$ and charge parameter $\chi_Q={q}/{\m}$. To simplify the computation, we arrange the infinite series expansion in the basis of homogeneous polynomials of $\chi$'s.

In $D=4$, the leading higher derivative corrections to Einstein-Maxwell theory are classified by their properties under parity transformations. For parity preserving 4-derivative interactions, we are able to obtain the perturbed solution up to ${\cal O}(\chi^7)$. Similar to the pure gravity case, 4-derivative interactions only add corrections to multipole moments that are actually nonzero at the leading order, namely, \(\{M_{2n+2}, \mathcal{S}_{2n+3}, \mathcal{Q}_{2n+2}, \mathcal{P}_{2n+1}\}\) for $n\ge 0$. Concerning the parity odd 4-derivative interactions, we obtain the perturbed solution up to ${\cal O}(\chi^8)$. We find that while these higher-derivative terms do not contribute to black hole thermodynamics, they affect the perturbative solutions, thereby modifying the black hole multipole moments. Notably, they contribute to multipole moments \(\{M_{2n+1}, \mathcal{S}_{2n}, \mathcal{Q}_{2n-1}, \mathcal{P}_{2n}\}\) for $n\ge1$ that all vanish at the leading order. This feature has significant observational implications. If they are observed, it would indicate the presence of parity odd higher-derivative corrections to General Relativity \cite{Fransen:2022jtw}.

As for future directions, we would like to push further to consider the next to next leading order higher derivative corrections and check if black hole multipole moments remain invariant under field redefinitions. The current method only allows us to carry out a case by case verification, after solving for the perturbed solution. For more general theories of gravity, in order to prove field redefinition invariance of black hole multipole moments, one may need to resort to a new formalism such as the one based on covariant phase space approach \cite{Compere:2017wrj}. Inspired by previous work \cite{Mayerson:2022ekj}, we propose generalizations of Geroch-Hansen formulae to broader gravity models, in which the generalized twist 1-form is derived using the covariant phase space approach. The results are presented in Appendix \ref{GH}. At this moment, we have not succeeded in applying these results to prove the field redefinition invariance of multipole moments and would like to pursue this problem in future study.

\section*{Acknowledgement}

We are grateful to Iosif Bena, Pablo Cano, Pengju Hu and Daniel Mayerson for useful discussions.
L.~Ma and Y.~Pang are supported by the National Key R\&D Program No.~2022YFE0134300 and the National Natural Science Foundation of China (NSFC) Grant No.~12175164, No.~12247103 and No.~12447138.
L.~Ma is also supported by Postdoctoral Fellowship Program of CPSF Grant No.~GZC20241211 and the China Postdoctoral Science Foundation under Grant No.~2024M762338. H. L\"{u} is supported in part by NSFC grants No.~11935009 and No.~12375052.

\appendix

\section{Black hole thermodynamics}
\label{thermo}
In this appendix, we shall verify our perturbative solution from a thermodynamic point of view. For asymptotically flat black hole like Kerr-Newman, there are at least two ways of computing the thermodynamic variables with leading higher derivative corrections. In the ordinary approach, one needs to first solve for the corrected solution and subsequently computes all the thermodynamic quantities using Wald procedure \cite{Wald:1993nt, Iyer:1994ys} or quasilocal formalism \cite{Brown:1992br}. The second approach was proposed by Reall and Santos \cite{Reall:2019sah}, which enables us to derive the leading higher derivative corrections to black hole thermodynamics using only the uncorrected solution.
The second approach has been rigorously tested in various problems related to the thermodynamics of rotating black holes \cite{Cheung:2019cwi, Ma:2022gtm, Mao:2023qxq,Ma:2024ynp}. Below, we will show that the black hole thermodynamics obtained from the perturbative solution \eqref{perturbed solution} using ordinary method agrees with that derived using the Reall-Santos method, demonstrating the legitimacy of our solution up to the approximation order. It is important to note that the parity odd 4-derivative interactions \eqref{odd} do not contribute black hole thermodynamics. Thus the results presented below encode corrections only from parity preserving 4-derivative terms.

\subsection{Black Hole thermodynamics from ordinary method}

For the perturbative solution \eqref{perturbed solution},
the black hole outer horizon is located at $r=r_h$, which satisfies
%%%%%%%%
\be
(g_{tt}g_{\varphi\varphi}-g_{t\varphi}^2)\Big|_{r=r_h}=0\quad \Longrightarrow\quad r_h=\mu+\mu  \sqrt{1-\chi ^2 (\chi _a^2+\chi _Q^2)}\ .
\ee
%%%%%%%
The angular velocity of the black hole is then given by
\bea
\Omega_H=-\Big(\frac{g_{t\varphi}}{g_{\varphi\varphi}}\Big|_{r=r_h}-\frac{g_{t\varphi}}{g_{\varphi\varphi}}\Big|_{r=\infty}\Big)\ .
\eea
Using the Killing vector $\xi=\partial_t+\Omega_H\partial_\varphi$ null at horizon, we compute the surface gravity $\k$ and temperature $T$ as
%%%%%%%
\be
\kappa^2=-\frac{g^{\mu\nu}\partial_\mu \xi^2\partial_\nu \xi^2}{4\xi^2}\Big|_{r=r_h},\quad T=\frac{\kappa}{2\pi}\ .
\ee
%%%%%%%
We then compute the entropy using Wald formula \cite{Wald:1993nt}
%%%%%%%%
\be
S=-\frac{1}{8}\int_{\cal B} d\O \frac{\partial {\cal L}}{\partial R_{\mu\nu\rho\sigma}}\epsilon_{\mu\nu}\epsilon_{\rho\sigma}\ ,
\ee
%%%%%%%
where ${\cal L}$ is the total Lagrangian and ${\cal B}$ is the bifurcation horizon.
The electric charge and potential are given by
%%%%%%%%%%%%%%%%%%
The electric charge is defined by the EOM
\be
Q_e=\frac{1}{16\pi}\int_{S^2}\star\mathcal{D}_{(2)},\quad \Phi_e=\xi^\mu A_\mu\Big|_{r=r_h}^{r=\infty}\ .
\label{electric charge}
\ee
%%%%%%%%%%%%%%%%%%
The energy and the angular momentum are obtained using Brown-York quasilocal stress tensor \cite{Brown:1992br}. After performing the redefinition of the integration constants \eqref{parameter redefinition}, the full set of thermodynamic quantities are summarized below.
%%%%%%%%%%%
\bea
M&=&\mu+\mathcal{O}(c_i^2,\chi^7)\ ,\quad Q_e=\frac{1}{2} \mu    \chi _Q+\mathcal{O}(c_i^2,\chi^7)\ ,\quad J=\mu ^2   \chi _a+\mathcal{O}(c_i^2,\chi^7)\ ,\nn\\
%%%
T&=&\frac{\sqrt{1- (\chi _a^2+\chi _Q^2)}}{2 \pi  \mu  \big(2 \sqrt{1- (\chi _a^2+\chi _Q^2)}- \chi _Q^2+2\big)}
+\delta T^{(e)}+\mathcal{O}(c_i^2,\chi^7)\ ,\nn\\
%%%
S&=&\pi  \mu ^2 \Big[\big(\sqrt{1- (\chi _a^2+\chi _Q^2)}+1\big)^2+ \chi _a^2\Big]+\delta S^{(e)}+\mathcal{O}(c_i^2,\chi^7)\ ,\nn\\
%%%
\Omega_H &=&\frac{  \chi _a}{\mu  \big(2 \sqrt{1- (\chi _a^2+\chi _Q^2)}- \chi _Q^2+2\big)}+\delta \Omega_H^{(e)}+\mathcal{O}(c_i^2,\chi^7)\ ,\nn\\
%%%
\Phi_e &=&\frac{2   \chi _Q (\sqrt{1- (\chi _a^2+\chi _Q^2)}+1)}{2 \sqrt{1- (\chi _a^2+\chi _Q^2)}- \chi _Q^2+2}+\delta \Phi_e^{(e)}+\mathcal{O}(c_i^2,\chi^7)\ .\label{thermo1}
\eea
where the correction terms take the form
%%%%%%%%%%
 \bea
 \delta T^{(e)}&=&-\frac{\alpha_2  \chi _Q^2}{32 \pi  \mu ^3}+\frac{ \chi _Q^2 }{640 \pi  \mu ^3}\Big[5 \alpha_2 (5 \chi _a^2-4 \chi _Q^2)+\alpha_0 \chi _Q^2\Big]+\frac{ \chi _Q^2 }{21504 \pi  \mu ^3}\Big[\chi _Q^2 \big(\alpha_0 (8 \chi _a^2+63 \chi _Q^2)\nn\\
 &&+128 (4 \alpha_1+\alpha_3) \chi _a^2\big)+2 \alpha_2 (776 \chi _a^2 \chi _Q^2+231 \chi _a^4-273 \chi _Q^4)\Big]\ ,\nn\\
 %%%
 \delta S^{(e)}&=&-\pi  \alpha_2  \chi _Q^2+\frac{\pi   \chi _Q^2}{40}  \Big[10 \alpha_2 \left(\chi _a^2-2 \chi _Q^2\right)+\alpha_0 \chi _Q^2\Big]+\frac{\pi   \chi _Q^2 }{3360}\Big[\chi _Q^2 \big(\alpha_0 (31 \chi _a^2+105 \chi _Q^2)\nn\\
 &&+160 (4 \alpha_1+\alpha_3) \chi _a^2\big)+2 \alpha_2 (550 \chi _a^2 \chi _Q^2+294 \chi _a^4-525 \chi _Q^4)\Big]\ ,\nn\\
 %%%
\delta \Omega_H^{(e)}&=&-\frac{\alpha_2  \chi _a \chi _Q^2}{8 \mu ^3}-\frac{ \chi _a \chi _Q^2 }{13440 \mu ^3}\Big[8 \alpha_2 (42 \chi _a^2+295 \chi _Q^2)+\big( 640 \alpha_1+160\alpha_3-11 \alpha_0\big) \chi _Q^2\Big]\nn\\
&&-\frac{ \chi _a \chi _Q^2 }{53760 \mu ^3}\Big[4 \alpha_2 (1520 \chi _a^2 \chi _Q^2+231 \chi _a^4+2555 \chi _Q^4)+\chi _Q^2 \big(32 (4 \alpha_1+\alpha_3) (13 \chi _a^2+35 \chi _Q^2)\nn\\
&&+\alpha_0 (201 \chi _a^2-35 \chi _Q^2)\big)\Big]\ ,\nn\\
%%%
\delta \Phi_e^{(e)}&=&\frac{\alpha_2  \chi _Q}{2 \mu ^2}-\frac{ \chi _Q }{40 \mu ^2}\Big[10 \alpha_2 (\chi _a^2-\chi _Q^2)+\alpha_0 \chi _Q^2\Big]-\frac{ \chi _Q }{6720 \mu ^2}\Big[\chi _Q^2 \big(\alpha_0 (20 \chi _a^2+231 \chi _Q^2)\nn\\
&&+320 (4 \alpha_1+\alpha_3) \chi _a^2\big)+2 \alpha_2 (1100 \chi _a^2 \chi _Q^2+399 \chi _a^4-420 \chi _Q^4)\Big]-\frac{ \chi _Q }{26880 \mu ^2}\Big[\chi _Q^2 \big( (445 \chi _a^2 \chi _Q^2\nn\\
&&+210 \chi _a^4+1008 \chi _Q^4)\alpha_0+32 (4 \alpha_1+\alpha_3)  (28 \chi _a^2+65 \chi _Q^2)\chi _a^2\big)+4  (2555 \chi _a^4 \chi _Q^2+2630 \chi _a^2 \chi _Q^4\nn\\
&&+483 \chi _a^6-420 \chi _Q^6)\alpha_2\Big]\ .\label{thermo2}
 \eea
%%%%%%%%%%%%%%%%%%%%%%%%%%%%%
In the expressions above, the leading terms in various thermodynamic quantities are exact in $\chi$. The precision level of the corrections is inherited from that of the perturbative solution which is at ${\cal O}(\chi^7)$.
%%%%%%%%%%%%%%%%%%%%%%%%%%%%%

\subsection{Black Hole thermodynamics from Reall-Santos method}

We now apply the Reall-Santos method \cite{Reall:2019sah} method to compute the same set of thermodynamic quantities. In the spirit of \cite{Reall:2019sah}, we simply plug the uncorrected solution into the total action with 4-derivative terms and integrate from the outer horizon to infinity. The resulting Euclidean action is a function of the uncorrected temperature $T_0$, angular velocity $\O_{H,0}$, and electric potential $\Phi_{e,0}$ whose expressions can be seen in \eqref{thermo1}. It is also straightforward to see that the parity odd terms in \eqref{odd} vanish on the purely electric black hole solution. Then similar to the results obtained using ordinary method, only parity even 4-derivative terms contribute to the thermodynamic quantities. The total Euclidean action takes the form
%%%%
\be
I_E(T_0,\O_{H,0},\Phi_{e,0})=T^{-1}_0G(T_0,\Omega_{H,0},\Phi_{e,0})\ ,\quad
G=G_0+\delta G^{(e)}\ ,
\ee
%%%%
where the leading order Gibbs free energy and its correction are given by
%%%%%
\bea
&&G_0=\frac{r_h^2+a^2}{4 r_h}+\frac{Q^2}{4 r_h}\left(1-\frac{2 r_h^2}{r_h^2+a^2}\right),\nn\\
%%%
&&\delta G^{(e)}=-\frac{Q^4(\alpha_0+8 \alpha_1+12 \alpha_2+2 \alpha_3)}{128 a^4 r_h^3}\Big(\frac{3 a^4+2 a^2 r_h^2+3 r_h^4}{r_h^2+a^2}+\frac{3 \left(a^4-r_h^4\right) \tan ^{-1}\Big(\frac{a}{r_h}\Big)}{a r_h}\Big)\nn\\
&&-\frac{Q^4 r_h (3 a^4-10 a^2 r_h^2+3 r_h^4)}{60 (r_h^2+a^2)^5}(\alpha_0-8 \alpha_1-4 \alpha_2-2 \alpha_3)
-\frac{2 (Q^2-2 \mu  r_h) \big(Q^2 r_h-\mu  (r_h^2-a^2)\big)}{(r_h^2+a^2)^3}\alpha_1\nn\\
&&
-\frac{2 Q^2 \big(4 Q^2 r_h (r_h^2-a^2)-\mu  (a^4-10 a^2 r_h^2+5 r_h^4)\big)}{5 (r_h^2+a^2)^4}\alpha_2\ .
\label{free RS}
\eea
%%%%%%%%%
Other thermodynamic quantities are derived from the Gibbs free energy according to
%%%%%%%%
\bea
S&=&-\frac{\partial G(T_0,\Omega_{H,0},\Phi_{e,0})}{\partial T_0}\Big|_{(\Omega_{H,0},\Phi_{e,0})}\ ,\quad Q_e=-\frac{\partial G^{(e)}(T_0,\Omega_{H,0},\Phi_{e,0})}{\partial \Phi_{e,0}}\Big|_{(T_0,\Omega_{H,0})}\ ,
\nn\\
J&=&-\frac{\partial G^{(e)}(T_0,\Omega_{H,0},\Phi_{e,0})}{\partial \Omega_{H,0}}\Big|_{(T_0,\Phi_{e,0})}\ ,\quad
M=G+T_0S+\Omega_{H,0}J+\Phi_{e,0}Q_e\ .
\eea
%%%%%%%
By this way, we obtain all the thermodynamic quantities in a specific choice of the integration constants such that the forms of temperature, angular velocity and electric potential are not modified by the 4-derivative interactions. Instead, the conserved charges such as the mass, angular momentum and electric charge do receive corrections. To compare with the results obtained using ordinary method, we need to redefine the integration constants in terms of which the conserved charges remain the same form as in the 2-derivative theory. Readers interested in the thermodynamic quantities derived from the Reall-Santos approach are also referred to the accompanying Mathematica file. After performing the appropriate redefinition of integration constant, we find that
%%%%%%
\be
M=M_0+\mathcal{O}(c_i^2),\quad Q_e=Q_{e,0}+\mathcal{O}(c_i^2),\quad J=J_0+\mathcal{O}(c_i^2)\ ,
\ee
and
\be
T=T_0+\delta T^{(e)},\quad S=S_0+\delta S^{(e)},\quad \Phi_{e}=\Phi_{e,0}+\delta \Phi_{e}^{(e)},\quad \Omega_{H}=\Omega_{H,0}+\delta \Omega_{H}^{(e)}\ ,
\label{thermodynamic RS 1}
\ee
where the corrections term indeed agree with those in \eqref{thermo2}.

\section{Redefinitions of integration constants}
\label{redefin}

In \eqref{parameter redefinition}, we have performed a redefinition of the integration constants $\m$, $\chi_a$ and $\chi_Q$ so that in terms of the new constants, the mass, angular momentum and electric charge remain the same form as in the 2-derivative theory. Below we list $\d\m$, $\d\chi_a$ and $\chi_Q$ up to certain order in $\chi$'s
%%%%%%%%%%%%%%%%%%%%%
\bea
\delta\mu&=&\frac{1 }{35 \mu }\Big[35 \alpha_2 \chi _a^2+\left(2 c_2+32 c_7-\alpha_0+4 \left(6 \alpha_1+\alpha_2+\alpha_3\right)\right) \chi _Q^2\Big]+\frac{1 }{13860 \mu }\Big[-1386 \alpha_2 \chi _a^4\nn\\
&&+3 \chi _a^2 \left(-24 c_2-384 c_7-65 \alpha_0+1252 \alpha_1+3201 \alpha_2+337 \alpha_3\right) \chi _Q^2+44 (2 c_2+32 c_7+3 \alpha_0\nn\\
&&+4 \left(6 \alpha_1+\alpha_2+\alpha_3\right)) \chi _Q^4\Big]+\frac{1 }{720720 \mu }\Big[-6 \chi _a^4 (864 c_2+13824 c_7-2148 \alpha_0+8652 \alpha_1-883 \alpha_2\nn\\
&&+1299 \alpha_3) \chi _Q^2+\chi _a^2 \left(-9280 c_2-148480 c_7-5999 \alpha_0+149128 \alpha_1+339144 \alpha_2+46562 \alpha_3\right) \chi _Q^4\nn\\
&&-27027 \alpha_2 \chi _a^6-1040 \left(2 c_2+32 c_7-9 \alpha_0+4 \left(6 \alpha_1+\alpha_2+\alpha_3\right)\right) \chi _Q^6\Big]+{\cal O}(\chi^7)\ ,
\nn\\
%%%
%%%
\delta \chi_a&=&\frac{\alpha_2 \chi _a}{\mu ^2}-\frac{1 }{420 \mu ^2}\Big[\chi _a \left(24 c_2+384 c_7-5 \alpha_0+148 \alpha_1-43 \alpha_2+13 \alpha_3\right) \chi _Q^2+462 \alpha_2 \chi _a^3\Big]\nn\\
&&+\frac{1 }{55440 \mu ^2}\Big[3465 \alpha_2 \chi _a^5+6 \chi _a^3 \left(48 c_2+768 c_7+262 \alpha_0-2372 \alpha_1-5511 \alpha_2-641 \alpha_3\right) \chi _Q^2\nn\\
&&+22 \chi _a \left(-16 c_2-256 c_7-45 \alpha_0+228 \alpha_1+178 \alpha_2+73 \alpha_3\right) \chi _Q^4\Big]+\frac{ \chi _a }{5765760 \mu ^2}\Big[108108 \alpha_2 \chi _a^6\nn\\
&&+6 \chi _a^4 \left(6912 c_2+110592 c_7-9891 \alpha_0+91524 \alpha_1+17389 \alpha_2+15969 \alpha_3\right) \chi _Q^2\nn\\
&&+5 \chi _a^2 \left(14848 c_2+237568 c_7+19637 \alpha_0-158296 \alpha_1-378924 \alpha_2-54422 \alpha_3\right) \chi _Q^4\nn\\
&&+260 \left(64 c_2+1024 c_7-387 \alpha_0+2748 \alpha_1+1019 \alpha_2+623 \alpha_3\right) \chi _Q^6\Big]+{\cal O}(\chi^8)\ , \nn
\\
%%%
\delta \chi_Q&=&-\frac{1 }{35 \mu ^2}\Big[35 \alpha_2 \chi _a^2 \chi _Q+\left(2 c_2+32 c_7-\alpha_0+4 \left(6 \alpha_1+\alpha_2+\alpha_3\right)\right) \chi _Q^3\Big]+\frac{1}{13860 \mu ^2}\Big[1386 \alpha_2 \chi _a^4 \chi _Q\nn\\
&&+3 \chi _a^2 \left(24 c_2+384 c_7+65 \alpha_0-1252 \alpha_1-3201 \alpha_2-337 \alpha_3\right) \chi _Q^3-44 (2 c_2+32 c_7+3 \alpha_0\nn\\
&&+4 \left(6 \alpha_1+\alpha_2+\alpha_3\right)) \chi _Q^5\Big]
+\frac{1 }{720720 \mu ^2}\Big[1040 \left(2 c_2+32 c_7-9 \alpha_0+4 \left(6 \alpha_1+\alpha_2+\alpha_3\right)\right) \chi _Q^7\nn\\
&&+\chi _a^2 \left(9280 c_2+148480 c_7+5999 \alpha_0-149128 \alpha_1-339144 \alpha_2-46562 \alpha_3\right) \chi _Q^5+6 \chi _a^4 \big(864 c_2\nn\\
&&+13824 c_7-2148 \alpha_0+8652 \alpha_1-883 \alpha_2+1299 \alpha_3\big) \chi _Q^3+27027 \alpha_2 \chi _a^6 \chi _Q\Big]+{\cal O}(\chi^8)\ .
\label{parameter redefinition 2}
\eea

\section{Equations of motion}
\label{eom}

In this section, we derive the field equations from the extended Einstein-Maxwell theory  described by the Lagrangian by $\mathcal{L}(g^{{\m\n}},R_{\mu\nu\rho\sigma},\widetilde{F}_{\mu\nu},F_{\mu\nu})$ in $D=4$.  Let us first define
%%%%%
\be
\widetilde{\mathcal{M}}^{\mu\nu}=-2\frac{\delta\mathcal{L}}{\delta \widetilde{F}_{\mu\nu}}\ ,\quad \mathcal{M}^{\mu\nu}=-2\frac{\delta\mathcal{L}}{\delta F_{\mu\nu}}\ ,\quad P^{\m\n\r\l}=\frac{\partial {\cal L}}{\partial R_{\m\n\r\l}}\ .
\ee

Then equations of motion for the metric and $U(1)$ gauge field take the form
\bea
0=E_{g,\mu\nu}&:=&\frac{1}{\sqrt{-g}}\frac{\d{(\sqrt{-g}\cal L)}}{\d g^{\m\n}}=P_{(\mu}^{\ \ \alpha\beta\gamma}R_{\nu)\alpha\beta\gamma}-\frac{1}{2}g_{\mu\nu}\mathcal{L}+2\nabla^\sigma\nabla^\rho P_{(\mu|\sigma|\nu)\rho}-\frac12\widetilde{\mathcal{M}}_{(\mu}^{\ \ \alpha}\widetilde{F}_{\nu)\alpha}
\nn\\
&&-\frac12\mathcal{M}_{(\mu}^{\ \ \alpha}F_{\nu)\alpha}+\frac{1}{4}g_{\mu\nu}\widetilde{\mathcal{M}}^{\alpha\beta}\widetilde{F}_{\alpha\beta}+\frac12\epsilon_{\alpha\beta\rho(\mu}F_{\nu)}^{\ \ \rho}\widetilde{\mathcal{M}}^{\alpha\beta}\ ,\nn\\
%%%
0=E_A^\mu&:=&\frac{\d{\cal L}}{\d A_{\m}}=\nabla_\nu\,{\cal D}^{\nu\mu}\ ,\quad
{\cal D}^{\mu\nu}=\frac{1}{2}\epsilon^{\mu\nu\rho\sigma}\widetilde{\mathcal{M}}_{\rho\sigma} +\mathcal{M}^{\mu\nu}\ .
\label{EOM high}
\eea

Specific to the 4-derivative extension of Einstein-Maxwell theory considered here, we divide the field equations into the leading pieces and the corrections
\be
E_{g,\mu\nu}=E_{g,\mu\nu}^{(0)}+\Delta E_{g,\mu\nu},\quad E_A^\m=E_A^{(0)\m}+\Delta E_A^\m\ .
\ee
The effective energy momentum tensor and electric current defined in \eqref{effective tensor} are given by
\bea
\Delta T_{\mu\nu}=-2\Delta E_{g,\mu\nu},\quad \Delta J^\m=-\Delta E_A^\m\ .
\eea

Analogously, the $P_{\m\n\r\s}$ and $\cal{M}_{\m\n}$ can also be separated into 2-and 4- derivative parts
\bea
P_{\mu\nu\rho\sigma}&=&P_{\mu\nu\rho\sigma}^{(0)}+\Delta P_{\mu\nu\rho\sigma}\ ,\quad
\mathcal{M}_{\mu\nu}=\mathcal{M}_{\mu\nu}^{(0)}+\Delta\mathcal{M}_{\mu\nu}\ ,
\nn\\
%%%
P_{\mu\nu\rho\sigma}^{(0)}&=&\frac{1}{2}\left(g_{\mu\rho}g_{\nu\sigma}-g_{\mu\sigma}g_{\nu\rho}
\right)\ ,\quad \mathcal{M}_{\mu\nu}^{(0)}=F_{\mu\nu}\ .
\eea
On the other hand, $\widetilde{M}_{\m\n}$ receive contributions only from parity odd 4-derivative terms. Corresponding to our parametrization of the 4-derivative actions \eqref{even} and \eqref{odd}, we have
\bea
\Delta P_{\mu\nu\rho\sigma}&=&\sum_{i=1}^8c_iP_{\mu\nu\rho\sigma}^{(e,i)}+\sum_{i=1}^4d_iP_{\mu\nu\rho\sigma}^{(o,i)},\nn\\
%%%
P_{\mu\nu\rho\sigma}^{(e,1)}&=&R\left(g_{\mu\rho}g_{\nu\sigma}-g_{\mu\sigma}g_{\nu\rho}
\right)\ ,\quad
P_{\mu\nu\rho\sigma}^{(e,2)}=\frac{1}{2}\left(g_{\nu\sigma}R_{\mu\rho}-g_{\nu\rho}R_{\mu\sigma}-g_{\mu\sigma}R_{\nu\rho}
+g_{\mu\rho}R_{\nu\sigma}
\right)\ ,\nn\\
%%%
P_{\mu\nu\rho\sigma}^{(e,3)}&=&2R_{\mu\nu\rho\sigma},\quad P_{\mu\nu\rho\sigma}^{(e,4)}=\frac{1}{2}\left(g_{\mu\rho}g_{\nu\sigma}-g_{\mu\sigma}g_{\nu\rho}
\right)F^2\ ,\nn\\
%%%
P_{\mu\nu\rho\sigma}^{(e,5)}&=&g_{\nu\sigma}F^2_{\mu\rho}-g_{\nu\rho}F^2_{\mu\sigma}-g_{\mu\sigma}F^2_{\nu\rho}
+g_{\mu\rho}F^2_{\nu\sigma},\quad P_{\mu\nu\rho\sigma}^{(e,6)}=F_{\mu\nu}F_{\rho\sigma},\quad P_{\mu\nu\rho\sigma}^{(e,7,8)}=0\ ,\nn\\
%%%
%%%
P^{(o,1)}_{\mu\nu\rho\sigma}&=&-\frac{1}{4}g_{\nu[\rho}\widetilde{F}_{\sigma]\mu}^2+\frac{1}{4}g_{\mu[\rho}\widetilde{F}_{\sigma]\nu}^2-\frac{1}{4}g_{\sigma[\mu}\widetilde{F}_{\nu]\rho}^2+\frac{1}{4}g_{\rho[\mu}\widetilde{F}_{\nu]\sigma}^2,\nn\\
%%%
P^{(o,2)}_{\mu\nu\rho\sigma}&=&\frac{1}{2}F_{\rho\sigma}\widetilde{F}_{\mu\nu}+\frac{1}{2}F_{\mu\nu}\widetilde{F}_{\rho\sigma},\quad P^{(o,3)}_{\mu\nu\rho\sigma}=0,\quad P^{(o,4)}_{\mu\nu\rho\sigma}=0\ .
\eea
and
%%%%%%%%
\bea
%%%
\Delta\mathcal{M}_{\mu\nu}&=&\sum_{i=1}^8c_i\mathcal{M}_{\mu\nu}^{(e,i)},\quad \widetilde{\mathcal{M}}_{\mu\nu}=\sum_{i=1}^4d_i\widetilde{\mathcal{M}}_{\mu\nu}^{(o,i)}\ ,\cr
%%%
%%%
\mathcal{M}_{\mu\nu}^{(e,1)}&=&0,\quad \mathcal{M}_{\mu\nu}^{(e,2)}=0,\quad \mathcal{M}_{\mu\nu}^{(e,3)}=0,\quad
\mathcal{M}_{\mu\nu}^{(e,4)}=-4F_{\mu\nu}R,\quad \mathcal{M}_{\mu\nu}^{(e,5)}=-4F_{[\mu}^{\ \rho}R_{\nu]\rho}\ ,\nn\\
%%%
\mathcal{M}_{\mu\nu}^{(e,6)}&=&-4R_{\mu\nu\rho\sigma}F^{\rho\sigma}\ ,\quad \mathcal{M}_{\mu\nu}^{(e,7)}=-4F_{\mu\nu}F^2\ ,\quad
\mathcal{M}_{\mu\nu}^{(e,8)}=-8F_\mu^{\ \rho}F_\nu^{\ \sigma}F_{\rho\sigma}\ ,\nn\\
%%%
%%%
\mathcal{M}_{\mu\nu}^{(o,1)}&=&-2\widetilde{F}_{[ \mu}^{\ \ \alpha}R_{\nu]\alpha},\quad \mathcal{M}_{\mu\nu}^{(o,2)}=-2R_{\mu\nu\rho\sigma}\widetilde{F}^{\rho\sigma},\quad \mathcal{M}_{\mu\nu}^{(o,3)}=-4\widetilde{F}^2F_{\mu\nu}-2F^2\widetilde{F}_{\mu\nu},\nn\\
%%%
\mathcal{M}_{\mu\nu}^{(o,4)}&=&-2F_{\mu\rho}F_{\nu\sigma}\widetilde{F}^{\rho\sigma}+2F^{\rho\sigma}F_{\nu\rho}\widetilde{F}_{\mu\sigma}-2F^{\rho\sigma}F_{\mu\rho}\widetilde{F}_{\nu\sigma},\nn\\
%%%
%%%
\widetilde{\mathcal{M}}_{\mu\nu}^{(o,1)}&=&-2F_{[ \mu}^{\ \ \alpha}R_{\nu]\alpha},\quad \widetilde{\mathcal{M}}_{\mu\nu}^{(o,2)}=-2R_{\mu\nu\rho\sigma}F^{\rho\sigma},\quad \widetilde{\mathcal{M}}_{\mu\nu}^{(o,3)}=-2F^2F_{\mu\nu}\ ,\nn\\
\widetilde{\mathcal{M}}_{\mu\nu}^{(o,4)}&=&-2F_{\mu\rho}F_{\nu\sigma}F^{\rho\sigma}\ ,
\eea
where we introduced the notation
\bea
\widetilde{F}^2_{\mu\nu}=F_{\mu\rho}\widetilde{F}_\nu^{\ \rho},\quad \widetilde{F}^2=F_{\mu\nu}\widetilde{F}^{\mu\nu}.\label{M tensor}
\eea

\section{Geroch-Hansen method for general theories of gravity}
\label{GH}

The gravitational multipole moments for asymptotically flat spacetimes in $D=4$ were defined by Geroch \cite{Geroch:1970cd} and Hansen \cite{Hansen:1974zz}. Utilizing the timelike Killing vector $\xi=\partial_t$, they constructed two scalars which encode the information about gravitational multipole moments. The first scalar is simply $\lambda=\xi^2=g^{tt}$. The second scalar arises from the twist 1-form of $\xi$ defined as
\be
\omega_{(1)}=i_{\xi}*d\xi\ ,
\label{twist one-form}
\ee
which has the property that
\be
d\omega_{(1)}=-\epsilon_{\mu\nu\rho\sigma}\xi^\rho R^\sigma_{\ \lambda}\xi^\lambda dx^{\m}\ .
\ee
Thus for Ricci-flat spacetimes, such as Kerr black hole, we can define the second scalar $\omega$ from the closure of $\o_{(1)}$
\be
\omega_{\m}=\partial_{\m}\o\ .
\ee
One can now combine $\lambda$ and $\o$ into two new scalars
\be
\Phi_M=\frac{1}{4\lambda}(\lambda^2+\omega^2-1),\qquad \Phi_J=\frac{\omega}{2\lambda}\ ,
\label{phimj}
\ee
which play the role of generating functions for the mass and current multipole moments respectively. Moreover, for Ricci-flat spacetimes, the equivalence between the Geroch-Hansen formalism and Thorne ACMC formalism was proven by \cite{Gursel}.

In \cite{Mayerson:2022ekj}, the author extended the Geroch-Hansen method to Einstein gravity with matter couplings in $D=4$. In particular, a closed 1-form was identified for a class of $N=2$ supergravity models, which generalizes the twist 1-form mentioned above. Below  we will use covariant phase space approach \cite{Wald:1993nt,Iyer:1994ys} to extend the result of \cite{Mayerson:2022ekj} to more general theories of gravity with higher derivative corrections.

From the above definition, we see that the key point of the Geroch-Hansen method is the construction of a closed 1-form $\omega_{(1)}$ from the Killing vector $\xi$, which further defines the second scalar \(\omega\). We start from a general matter coupled theory of gravity described by a diffeomorphism invariant Lagrangian 4-form ${\mathbf L}(\Phi)$,
where $\Phi$ is a shorthand notation for all the fields involved.
Consider an infinitesimal coordinate transformation generated by a local vector field $\eta$
\bea
\delta_\eta {\mathbf L}[\Phi]=\mathbf{E}_\Phi\delta_\eta\Phi+d\mathbf{\Theta}(\Phi,\delta_\eta\Phi)\ .
\eea
Using the on-shell condition $\mathbf{E}_\Phi=0$ and Cartan magic formula $\delta_\eta=di_\eta+i_\eta d$, we can define a conserved Noether current $\mathbf{J}_\eta$ whose closure implies the existence of the Noether charge
$\mathbf{Q}_\eta$
%%%%
\bea
\mathbf{J}_\eta &=&\mathbf{\Theta}(\Phi,\delta_\eta\Phi)-i_\eta\mathbf{L}[\Phi],
\nn\\
d\mathbf{J}_\eta&=&0\quad\Longrightarrow \quad\mathbf{J}_\eta=d{\mathbf{Q}_{\eta}}\ .
\eea
%%%%

Choosing $\eta$ to be the Killing vector $\xi$, we have $\delta_\xi\Phi=0$ and thus
$\mathbf{\Theta}(\Phi,\delta_\xi\Phi)=0$. Consequently, onshell we have
%%%%%%%%%%
\be
di_\xi\mathbf{L}[\Phi]=0 \quad \Longrightarrow\quad
i_\xi\mathbf{L}[\Phi]=-d\Omega.
\ee
The definition of $\mathbf{J}_\xi$ implies that we can form a closed 2-form from
$\mathbf{Q}_{\xi}$ and $\Omega$ as
%%%
\be
d(\mathbf{Q}_\xi-\Omega)=0 \quad \Longrightarrow \quad \widetilde{\mathbf{Q}}_\xi=\mathbf{Q}_\xi-\Omega+d\mathbf{Y}\ ,
\ee
%%%%
where $\mathbf{Y}$ is ambiguity that one can play with.
The generalized twist 1-form \eqref{twist one-form} and its potential are then given by
\bea
\omega_{(1)}=i_\xi\widetilde{\mathbf{Q}}_\xi=d\omega.\label{improve scalar omega}
\eea
For Einstein-Maxwell theory and the Einstein-Maxwell-dilaton theory, the expression of $\o_{(1)}$ is given in \cite{Mayerson:2022ekj,Liu:2022wku}.
The author has pointed out, one can choose the ambiguous term $\mathbf{Y}$ to ensure that  the matter contributions to $\o$ will not affect the multipole moments. This choice of $\mathbf{Y}$ is also consistent with the requirement of ACMC coordinate system. So far, since the scalar $\l$ appearing in \eqref{phimj} does not have a direct origin from the Lagrangian, we have not been able to use these results to prove that multipole moments are invariant under field redefinitions.

%\bibliographystyle{utphys}
%\bibliography{ref}

\end{document}